\documentclass[twocolumn,
		spanish,
		aps,
		ams,
		showkeys,
		superscriptaddress]{revtex4-1}

\usepackage{amsmath}
\usepackage{amssymb}
\usepackage{graphicx}
\usepackage{color}
\usepackage[draft=false]{hyperref}

\begin{document}

\title{Behavior of Phantom Scalar Fields near Black Holes}
\author{F. D. Lora-Clavijo, J. A. Gonz\'alez, F. S. Guzm\'an}
\affiliation{Instituto de F\'{\i}sica y Matem\'{a}ticas, Universidad
              Michoacana de San Nicol\'as de Hidalgo. Edificio C-3, Cd.
              Universitaria, 58040 Morelia, Michoac\'{a}n,
              M\'{e}xico.}
\date{\today}

\begin{abstract}
We present the accretion of a phantom scalar field into a black hole for various scalar field potentials in the full non-linear regime. Our results are based on the use 
of numerical methods and show 
that for all the cases studied the black hole's apparent horizon mass decreases. We explore a particular subset of the parameter space and from our 
results we conclude 
that this is a very efficient black hole 
shrinking process 
because the time scales of the area reduction of the horizon are short. 
We show that the radial equation of state of the scalar field depends strongly on the space and time, with the condition $\omega = p/\rho>-1$, as opposed to a 
phantom fluid at cosmic scales that allows $\omega < -1$.
\end{abstract}

\maketitle






\section{Introduction}

In cosmological models at present time the universe is assumed to have various ingredients 
of exotic nature like the dark energy candidates, among which we find the phantom 
scalar field  \cite{Corasaniti}. The reason is that supernovae Ia data allow an equation of 
state of the dark energy component with $\omega = 
p/\rho <-1$, where $p$ and $\rho$ are the pressure and the energy 
density of the fluid, which is a condition that a phantom scalar field satisfies at cosmic 
scales.
This is a reason to start up an exploration of the consequences such type of scalar 
field might have at local scales, for instance on black holes. On the other hand, the behavior 
of the area of the horizon when the black holes is accreting matter is a very important 
property of black hole physics, because the accretion of such exotic material may impose 
important restrictions on the mass of black nowadays black hole candidates.

In order to study this process in the full non-linear regime we start up with a model coupling 
the phantom scalar field and gravity. The model assumes the Lagrangian density of the phantom 
scalar field is given by 

\begin{equation}
{\cal L} = R + \frac{1}{2}g_{\mu\nu}\partial^{\mu}\phi \partial^{\nu} \phi - V(\phi), 
\label{eq:lagrangian}
\end{equation}

\noindent where $R$ is the Ricci scalar of the space-time, $g_{\mu\nu}$ is the 
space-time metric, $\phi$ is a scalar field and $V(\phi)$ is the potential of the field.
The property defining a 
phantom scalar field is that the relative sign between the Ricci scalar 
and the kinetic term are the same. When the action constructed 
with such a Lagrangian density is varied with respect to the metric, the 
arising Einstein's equations are related to a stress energy-tensor that 
violates the null energy condition, that is $T_{\mu\nu}k^{\mu}k^{\nu} < 0$, 
where $k^{\mu}$ is a null vector. The immediate implication of this 
violation is the violation of the weak energy condition too, 
which in turn implies that observers following time-like trajectories 
might measure negative energy densities. Although this property is at odds with 
the nowadays physics observed in laboratories, there are cosmological observations
indicating the presence of such kind of matter\cite{Corasaniti}. In this paper
we explore the implications of this at astrophysical scales.

The fact that the scalar field violates the null energy condition 
motivates the study of possible unusual implications in astrophysical 
scenarios, because the area increasing theorem does not apply in this case. 
For instance, using exact solutions corresponding to stationary accretion of a test  
phantom fluid, it was found that the 
mass of black holes decreases in a phantom energy dominated universe 
approaching the big rip \cite{Babichev}. 
It has also been studied recently the behavior of a black hole apparent 
horizon (AH) in a FRW background, and the conditions under which a naked 
singularity can be formed due to the coincidence of the AH of the black 
hole and the cosmic horizon \cite{Faraoni}; in such case 
the authors consider the effects of the back reaction of the scalar field 
on the space-time metric and indicate that cosmic censorship does not 
only forbid the existence of naked singularities but also the existence 
of a phantom field.
Assuming that the null energy condition is satisfied the area of the horizon only increase. 
However, if this condition is violated, the area of the event horizon decreases and the black 
hole shrinks as shown recently in the full non-linear regime using numerical relativity 
in \cite{Guzman}.
The present paper is a detailed follow up of previous one, in which we also explore the effects 
of the scalar field potential on the accretion rates and final state of the black hole.

Two important items are presented in this paper: i) the relation $p/\rho < -1$ is not fulfilled 
at local scales by the scalar field (at least near to a black hole) although the null energy 
condition is not satisfied and ii) the accretion of such scalar field reduces the area of a black 
hole at similar rates for different types of potentials driving the scalar field.

This paper is organized as follows. In section \ref{sec:cauchy} we 
describe the 3+1 decomposition of the space-time, the evolution system of 
equations driving the evolution of the geometry and the construction of the initial data. In section 
\ref{sec:results} we present the results obtained. Finally in section 
\ref{sec:conclusions} we draw some conclusions and comments.


\section{The system of equations}
\label{sec:cauchy}

 We formulate Einstein's field equations coupled to the scalar field in such a way that these can be integrated numerically. In this model, we will use spherical 
symmetry and use geometrized units for which the speed of light and Newton's constant are equal to one.

In order to the study the dynamics of a spherically symmetric space-time, we write the general metric for the coordinate system 
$(t,r,\theta,\vartheta)$ in the following form
\begin{eqnarray}
ds^2 &=& -( \alpha^2 - \beta^r \beta^r \frac{g_{rr}}{\chi} )dt^2
         + 2\beta_r dtdr  \nonumber\\
        &+& \frac{1}{\chi} [g_{rr} dr^2 + g_{\theta\theta}
        (d\theta^2 + \sin^2 \theta d\phi^2)], \label{eq:metric}
\end{eqnarray}
\noindent where $\chi$ acts as a conformal factor relating this 
metric to a space-like flat metric, $\beta^r$ is the only non-zero 
component of the shift vector and $\alpha$ is the lapse function as in 
\cite{Brown}.

We solve the evolution Einstein's field equations using the Generalized 
BSSN evolution formulation of the 3+1 decomposition of General 
Relativity described in \cite{Brown} as opposed to previous successful 
analyzes refereed to the accretion of scalar field using 
Eddington-Finkelstein like coordinates under the ADM formulation \cite{Thornburg}.


\subsection{Evolution}
\label{subsec:evolution}

For the construction of the space-time we carry out a Cauchy-type evolution of initial 
data based on the 3+1 decomposition of the space-time. Within such decomposition 
there are various formulations of the evolution equations with different hyperbolic properties. 
Among such formulations the most popular nowadays is the so called BSSN 
formulation \cite{BSSN} which helped at solving the problem of the binary black hole collision system 
recently \cite{NASA,Brownsville} using the punctures technique that allowed the adequate treatment of the 
black hole singularity \cite{Bruegmann}.
The BSSN formulation assumes that the conformal metric has determinant equal to one, but in spherical
coordinates the flat metric the determinant is different to one. This issue is addressed 
relaxing this condition over the determinant, obtaining the Generalized BSSN equations (GBSSN) 
\cite{Brown2}, which is the formulation we use in our simulations. 
The GBSSN system of equations in a spherically symmetric space-time reduce to a set of six 
equations for the independent dynamical variables $\chi$, $g_{rr}$, $g_{\theta\theta}$, the non-zero 
trace-free part of the extrinsic curvature $A_{rr}$, the trace of the extrinsic curvature $K$ 
and the contracted Christoffel non-zero symbol (conformal connection function) $\Gamma^{r}$. 
The explicit expressions in the presence of matter are:

\begin{widetext}

\begin{eqnarray*}
&& \partial_t g_{rr} = -2A_{rr}\alpha - \frac{v\beta^{r}g_{rr}'}{3} + \beta^{r}g_{rr}' 
- \frac{2g_{rr}v\beta^{r}g_{\theta \theta}'}{3g_{\theta \theta}} + 2g_{rr}\beta^{r'} 
- \frac{2g_{rr}v\beta^{r'}}{3}, \nonumber\\ 
&& \partial_t g_{\theta \theta} = \frac{A_{rr}g_{\theta \theta}\alpha}{g_{rr}} 
- \frac{g_{\theta \theta}v\beta^{r}g_{rr}'}{3g_{rr}} - \frac{2v\beta^{r}g_{\theta \theta}'}{3} 
+ \beta^{r}g_{\theta \theta}' - \frac{2g_{\theta \theta}v\beta^{r'}}{3}, \nonumber\\ 
&& \partial_t \chi = \frac{2K\alpha \chi}{3} - \frac{v\beta^{r}g_{rr}'\chi}{3g_{rr}} 
- \frac{2v\beta^{r}g_{\theta \theta}'\chi}{3g_{\theta \theta}} - \frac{2v\beta^{r'}\chi}{3} 
+ \beta^{r}\chi', \nonumber\\
&& \partial_t A_{rr} = -\frac{2\alpha A^{2}_{rr}}{g_{rr}} + K\alpha A_{rr} 
- \frac{v\beta^{r}g_{rr}'A_{rr}}{3g_{rr}}	- \frac{2v\beta^{r}g_{\theta \theta}'A_{rr}}{3g_{\theta \theta}} 
- \frac{2v\beta^{r'}A_{rr}}{3} + 2\beta^{r'}A_{rr} + \frac{2\alpha \chi (g_{rr}')^2}{3g^{2}_{rr}} 
- \frac{\alpha \chi (g_{\theta \theta}')^2}{3g^{2}_{\theta \theta}}  \nonumber\\
&& \qquad \qquad -\frac{\alpha (\chi')^2}{6\chi} - \frac{2g_{rr}\alpha \chi}{3g_{\theta \theta}} 
+ \beta^{r}A_{rr}' + \frac{2g_{rr}\alpha \chi \Gamma^{r'}}{3} 
- \frac{\alpha \chi g_{rr}' g_{\theta \theta}'}{2g_{rr}g_{\theta \theta}} 
+ \frac{\chi g_{rr}' \alpha'}{3g_{rr}} + \frac{\chi g{\theta \theta}' \alpha'}{3g_{\theta \theta}} 
- \frac{\alpha g_{rr}' \chi'}{6g_{rr}} - \frac{\alpha g_{\theta \theta}' \chi'}{6g_{\theta \theta}} \nonumber\\ 
&& \qquad \qquad - \frac{2\alpha' \chi'}{3} - \frac{\alpha \chi g_{rr}''}{3g_{rr}} 
+ \frac{\alpha \chi g_{\theta \theta}''}{3g_{\theta \theta}} - \frac{2\chi\alpha''}{3} 
+ \frac{\alpha \chi''}{3} - \chi \alpha M_{rr},\nonumber\\
&& \partial_t K = \frac{3\alpha A^{2}_{rr}}{2 g^{2}_{rr}} + \frac{K^{2}\alpha}{3} 
+ \beta^{r}K' + \frac{\chi g_{rr}' \alpha'}{2g^{2}_{rr}} 
- \frac{\chi g_{\theta \theta}' \alpha'}{g_{rr}g_{\theta \theta}} + \frac{\alpha' \chi'}{2g_{rr}} 
- \frac{\chi \alpha''}{g_{rr}} + \frac{\alpha}{2}(\rho + S),\nonumber\\
&& \partial_t \Gamma^{r} = -\frac{v\beta^{r} (g_{\theta \theta}')^2}{g_{rr}g^{2}_{\theta \theta}} 
+ \frac{A_{rr} \alpha g_{\theta \theta}'}{g^{2}_{rr}g_{\theta \theta}} 
- \frac{v\beta^{r'} g_{\theta \theta}'}{3g_{rr}g_{\theta \theta}} 
+ \frac{\beta^{r'} g_{\theta \theta}'}{g_{rr}g_{\theta \theta}} - \beta^{r}\Gamma^{r'} 
+ \frac{A_{rr}\alpha g_{rr}'}{g^{3}_{rr}} - \frac{4\alpha K'}{3g_{rr}} 
- \frac{2A_{rr}\alpha'}{g^{2}_{rr}} + \frac{vg_{rr}'\beta^{r'}}{2g^{2}_{rr}} \nonumber\\ 
&& \qquad\qquad - \frac{g_{rr}'\beta^{r'}}{2g^{2}_{rr}} - \frac{3A_{rr}\alpha\chi'}{g^{2}_{rr}\chi} 
+ \frac{v\beta^{r}g_{rr}''}{6g^{2}_{rr}} + \frac{v\beta^{r}g_{\theta\theta}''}{3g_{rr}g_{\theta\theta}} 
+ \frac{v\beta{r''}}{3g_{rr}} + \frac{\beta^{r''}}{g_{rr}} 
- \frac{2\alpha}{g_{rr}}S_{r}. \label{eq:GBBSN_evolution}
\end{eqnarray*}

\end{widetext}
\noindent where primes denote derivatives with respect to $r$ and the gauge parameter $v$ is 
such that for $v=0$ the coordinates are Eulerian, whereas for $v=1$ the coordinates are Lagrangian.

We have introduced the matter sources $\rho$, $S_{r}$, $M_{rr}$ and $S$ which represent the energy 
density, the momentum current density, the non-zero trace free part of the stress tensor and the 
trace of the stress tensor respectively. 
These quantities can be computed after projecting the stress-energy tensor onto the unit normal 
vector $n_{\alpha}$ to the space-like hyper-surfaces,

\begin{eqnarray}
\nonumber  \rho &=& n_{\alpha}n_{\beta}T^{\alpha\beta},\\ \nonumber
 S_{r} &=& -\gamma_{r\alpha}n_{\beta}T^{\alpha\beta}, \\ \nonumber
 M_{rr} &=& \gamma_{r\alpha}\gamma_{r\beta}T^{\alpha\beta}, \nonumber\\
 S &=& \gamma^{ij}S_{ij}.\nonumber
\end{eqnarray}

The constraints of the system are: the Hamiltonian constraint $H$, the momentum 
constraint $M_{i}$ and the constraint arising from the definition of the conformal connection 
functions $G^{i}$. In spherical symmetry they are:

\begin{subequations}\begin{eqnarray}
\nonumber && H = -\frac{3A^{2}_{rr}}{2g^{2}_{rr}} + \frac{2K^{2}}{3} - \frac{5(\chi')^2}{2\chi g_{rr}} 
+ \frac{2\chi''}{g_{rr}} + \frac{2\chi}{g_{\theta\theta}} 
- \frac{2\chi g_{\theta\theta}''}{g_{rr}g_{\theta\theta}}\\ 
&& \qquad + \frac{2\chi' g_{\theta\theta}'}{g_{rr}g_{\theta\theta}} 
+ \frac{\chi g_{rr}' g_{\theta\theta}'}{g^{2}_{rr}g_{\theta\theta}} - \frac{\chi' g_{rr}'}{g^{2}_{rr}} 
+ \frac{\chi (g_{\theta\theta}')^2}{2g_{rr}g^{2}_{\theta\theta}} - 2\rho, \\ \nonumber
&& M_{r} = \frac{A_{rr}'}{g_{rr}} - \frac{2K'}{3} - \frac{3A_{rr}\chi'}{2\chi g_{rr}} \\ 
&& \qquad \qquad \qquad  + \frac{3A_{rr}g_{\theta\theta}'}{2g_{rr}g{\theta\theta}} 
- \frac{A_{rr}g_{rr}'}{g^{2}_{rr}} - S_{r}, \\
&& G^{r} = -\frac{g_{rr}'}{2g^{2}_{rr}} + \Gamma_{r} + \frac{g_{\theta\theta}'}{g_{rr}g_{\theta\theta}}. 
\label{eq:constraint}
\end{eqnarray}\end{subequations}

In order to specify the gauge, we evolve the lapse according to the standard  
1+$\log$ slicing condition 

\begin{eqnarray}
\partial_t \alpha = \beta^a \partial_a\alpha - 2\alpha K, \label{eq:lapse}
\end{eqnarray}

\noindent or the related condition obtained by dropping the advection term $\beta^{a}\partial_{a}\alpha$.
For the shift vector we implemented the recipe for the $\Gamma$-driver condition

\begin{eqnarray}
\partial_t \beta^a &=& \frac{3}{4}B^{a} + \beta^c \partial_c	\beta^a,\nonumber\\
\partial_t B^a &=& \partial_t \Gamma^a + \beta^c\partial_c B^a - \beta^c \partial_c \Gamma^a - \eta B^a,
\label{eq:shift}
\end{eqnarray}

\noindent which helps to avoid the slice stretching near the horizon, which is known to kill the 
numerical evolution. In order to avoid instabilities near the origin (the puncture), we 
implement a sort of excision without excision \cite{excision} using a factor function on the source of the evolution 
equations of the form $(r/(1+r))^4$ from the coordinate origin out to $r_{ex}$ an appropriate fraction 
of the size of the apparent horizon radius such that $r_{ex}$ is smaller that the apparent horizon of 
the final black hole. Even though this function violates the constrains, the violations do not 
propagate out of the black hole horizon as shown by the convergence tests.
In all our simulations we use Eulerian coordinates, that is, we set $v=1$.

The evolution of the scalar field is driven by the Klein-Gordon equation

\begin{equation}
\Box \phi -\frac{dV}{d\phi}= \frac{1}{\sqrt{-g}}\partial_{\mu}[\sqrt{-g}g^{\mu\nu} \partial_{\nu}\phi]  -\frac{dV}{d\phi}=0. \label{eq:wave}
\end{equation}

\noindent In order to reduce this equation to a first order system as the geometric counterpart, we define two new variables $\pi=\partial_t \phi$ and $\psi=\partial_r \phi$, and (\ref{eq:wave}) reduces to the system

\begin{eqnarray}
\partial_t \pi &=& \partial_t \beta^r
   + (-\pi + \beta^r) \partial_t \ln
   \left( \frac{\sqrt{g_{rr}}g_{\theta\theta}}{\alpha \chi^{3/2}} 
\right)\nonumber\\
   &+& \partial_r \ln
   \left( \frac{\sqrt{g_{rr}}g_{\theta\theta}}{\alpha \chi^{3/2}} 
\right)
   \left[ \left(\frac{\alpha^2 \chi}{g_{rr}} - \beta^r \right)\psi
        + \beta^r \pi \right] \nonumber\\
   &+& \partial_r
   \left[ \left(\frac{\alpha^2 \chi}{g_{rr}} - \beta^r\right)\psi + 
\beta^2 \pi \right] + \alpha^2 \frac{dV}{d\phi}
\nonumber\\
\partial_t \psi &=& \partial_r \pi . \label{eq:sf_evolution}
\end{eqnarray}

\noindent which together with the evolution equations for the geometry and the  completes the system of equations to be solved.


\subsection{Initial data}
\label{subsec:initialdata}

In order to start up an evolution including the matter terms it is necessary to 
solve the constraints a the initial time slice, and afterwards Bianchi identities guarantee that such constraints are satisfied. We assume that the 
initial slice is time-symmetric, which implies that the shift vector $\beta^{r}$ and its time derivative are zero initially. In addition, time symmetry imply that all 
the components of the extrinsic curvature are zero initially. In this way, the momentum constraint is satisfied identically at initial time.
On the other hand, we use the pre-collapsed condition for the lapse $\alpha=(1+M/2r)^{-2}$. 

In order to provide a local nature to the scalar field we assume that $\phi$ has a Gaussian profile or is a train of Gaussians one next to the other, which implies 
initial data for the first order variable $\psi$, and the time symmetry provides the condition $\pi=0$. With this information about the matter source we solve the 
Hamiltonian constraint. Now, in order to achieve a smooth coordinate system we use the initial ansatz that the space-time metric has a form similar to that of a Schwarzschild black hole in isotropic coordinates. Thus we assume that the metric at initial time has the form

\begin{equation}
g_{rr} = 1, ~~~
g_{\theta \theta} = r^2, ~~~
\chi = \left(1 + \frac{M}{2r} + u\right)^{-4} 
\label{eq:metric_id}
\end{equation}

\noindent where $M$ is the mass of the apparent horizon and $u=u(r,t=0)$ 
is the function to be determined through the solution of the Hamiltonian 
constraint. The Hamiltonian constraint is thus reduced to the equation

\begin{eqnarray}
\nonumber && \partial_{rr} u = \frac{(\partial_{r} \phi)^2}{8} \left(1+\frac{M}{2r} + u \right) \\
	&&  -\frac{2 \partial_r u}{r} -\frac{V(\phi)}{4}\left(1+\frac{M}{2r} + u \right)^5, 
\label{eq:u}
\end{eqnarray}

\noindent which we solve using a fourth order Runge-kutta 
integrator. In the last equation (\ref{eq:u}), $V(\phi)$ is the self-interaction potential of the phantom scalar field.

We finally rescaled the coordinate $r$ in such way that $\chi\rightarrow 1$ when $r\rightarrow\infty$.
These initial data are used to start up a Cauchy type evolution using the 
GBSSN evolution equations (\ref{eq:GBBSN_evolution}), the gauge conditions (\ref{eq:lapse} - \ref{eq:shift}) and the scalar field equation (\ref{eq:sf_evolution}).


\subsection{Implementation and diagnostics}
\label{subsec:diagnostics}

The numerical method used to approximate the constraint and evolution equations is a fourth order finite differences approximation. We only perform the evolution on a finite domain with 
artificial boundaries at a finite value of $r$ where we implement  
radiative-type boundary conditions. 
The integration in time uses a method of lines with 
a fourth order accurate Runge-Kutta integrator. Throughout the 
evolution, we monitor the Hamiltonian, Momentum and $G^r$ constraints, so that we check 
that they converge to zero as resolution is increased with fourth order.

{\it Apparent horizons}. In order to track the radius, area and 
mass of the apparent horizon during the evolution in terms of the $3+1$ variables, we search for the  
marginally trapped surfaces (MTS) through the condition

\begin{equation}
\Theta = \nabla_i n^i + K_{ij} n^i n^j - K =0, \label{eq:MTS} 
\end{equation}

\noindent where we take $n^i$ to be the outward pointing unit vector normal to the horizon, $K_{ij}$ are the components of the extrinsic curvature and $K$ its trace of a space-like 
hypersurface on which one calculates the MTSs. The apparent horizon 
is the outermost MTS. In our coordinates equation (\ref{eq:MTS}) for the metric (\ref{eq:metric}) reads

\begin{equation}
\frac{\partial_{r}g_{\theta\theta}}{g_{\theta\theta}} - \frac{\partial_{r}\chi}{\chi} - 2\sqrt{\frac{g_{rr}}{\chi}}\left(\frac{K}{3} - \frac{A_{rr}}{2g_{rr}} \right) = 0. 
\label{eq:ah}
\end{equation}

\noindent In order to track the evolution of the apparent horizon we 
calculate $\Theta$ at every time step and locate the outermost zero of 
it at the coordinate radius $r_{AH}$. Then calculate the area of 
the corresponding 2-sphere $A_{AH}=4\pi R_{AH}^{2}$ and its mass $M_{AH} 
= R_{AH}/2$, where $R_{AH}=\sqrt{g_{\theta\theta / \chi}}$ is the areal radius evaluated at $r_{AH}$.

{\it Misner-Sharp mass function} This quantity is a space-dependent measure of the mass of the space-time. We use this function in order to compare the mass of the apparent horizon and the mass of the space-time, so that we can monitor the consistency of our simulations.
The Misner-Sharp mass function in our case is written as

\begin{eqnarray}
\nonumber &&  M_{MS} = \frac{R}{2}\left[1 + \frac{(\partial_{t}R)^2}{\alpha^2} -\frac{2\beta^{r}(\partial_{t}R)(\partial_{r}R)}{\alpha^2} \right. \\ 
&&  \left. \qquad \qquad  - \left( \frac{\chi}{g_{rr}} - \frac{(\beta^{r})^2}{\alpha^2}\right)(\partial_{r}R)^2 \right], 
\end{eqnarray}

\noindent where $R=\sqrt{g_{\theta\theta} / \chi}$.

The Misner-Sharp mass reduces to the Arnowitt-Deser-Misner (ADM) at space-like infinity, 

\begin{equation}
 M_{ADM} = \lim_{r\rightarrow\infty}M_{MS},
\end{equation}

\noindent so we can estimate the mass of the space-time at a finite radius at any time during the evolution. 
On the other hand, the Misner-Sharp mass function reduces to the Bondi-Sachs mass at future null infinity.

{\it Event horizon}. Aside of the apparent horizon which is a gauge dependent 2-surface, we track a bundle of radial outgoing null rays in order to approximately locate 
the event horizon during the evolution as the 
3-surface from which outgoing null rays diverge when launched toward the future. For this we solve the geodesic equation for radial null rays on the 
fly during the evolution, and fine-tune their location initially so that both, outgoing null rays escaping toward future null infinity and outgoing null rays that 
reach the singularity converge to the same surface for as much time as possible during the evolution, after which they spread. Such surface is the approximate location 
of the event horizon.


\section{Results}
\label{sec:results}

\begin{figure*}[ht]
\includegraphics[width=7cm]{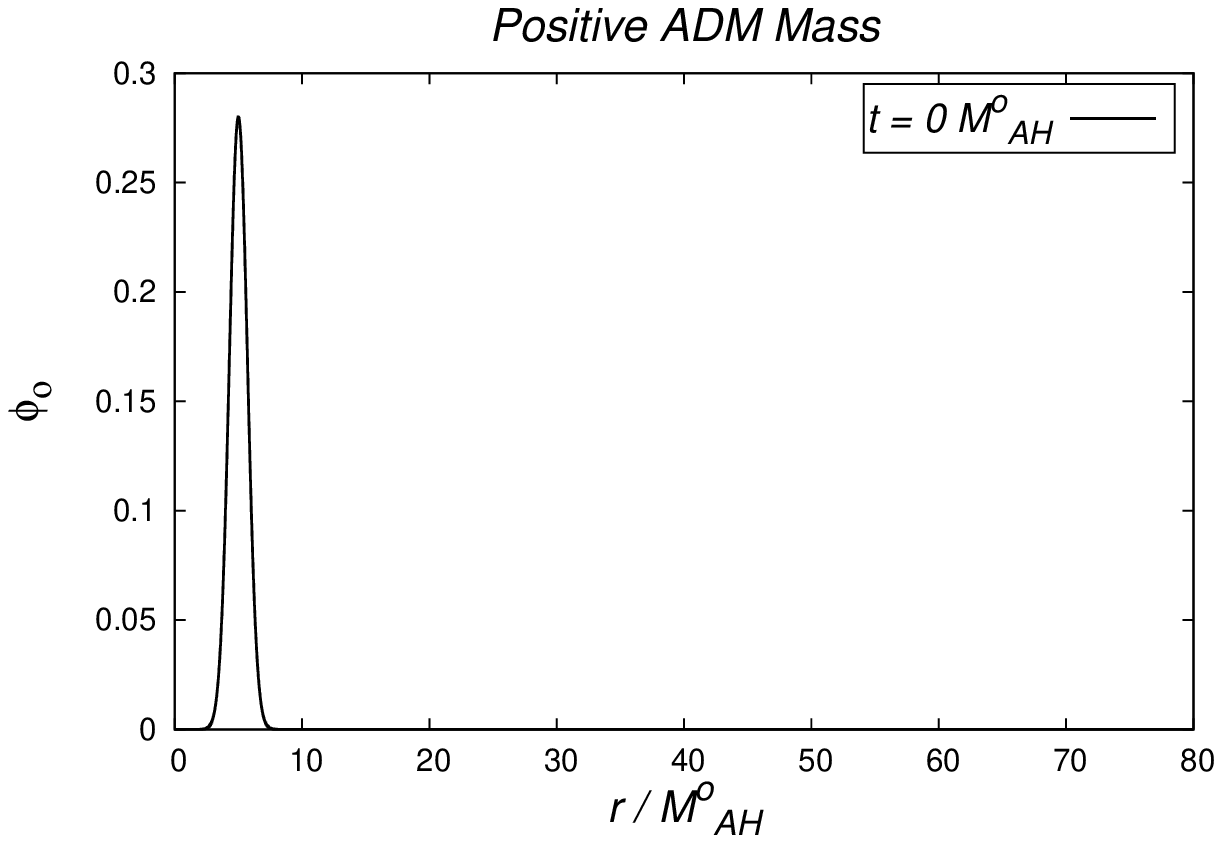}
\includegraphics[width=7cm]{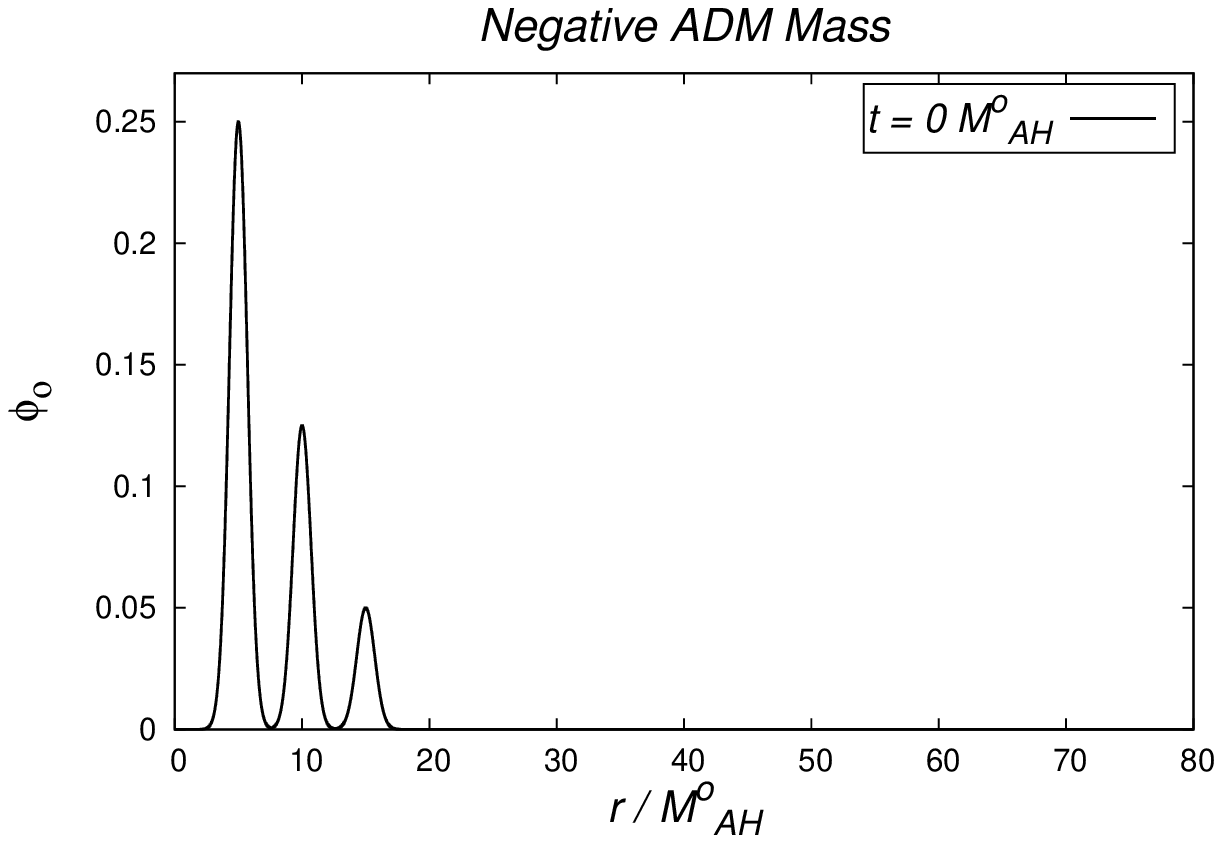}
\includegraphics[width=7cm]{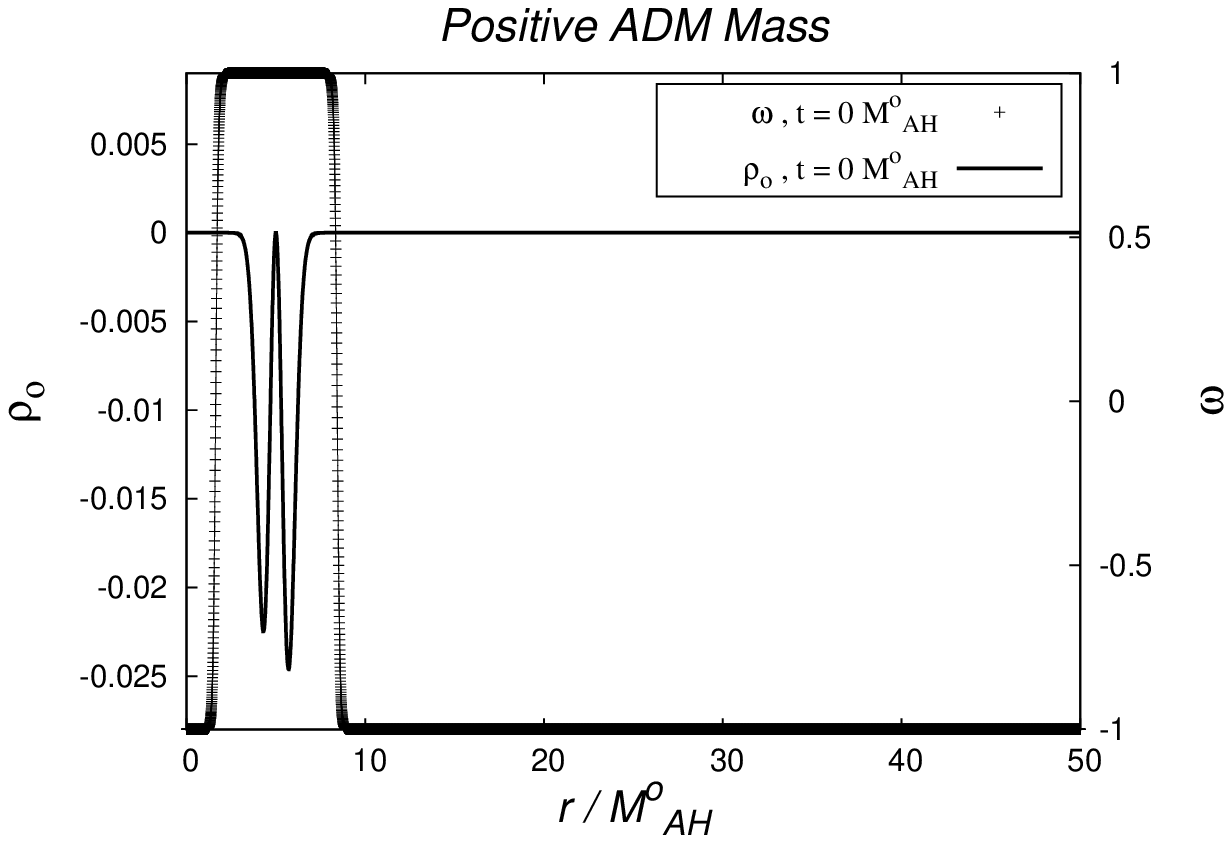}
\includegraphics[width=7cm]{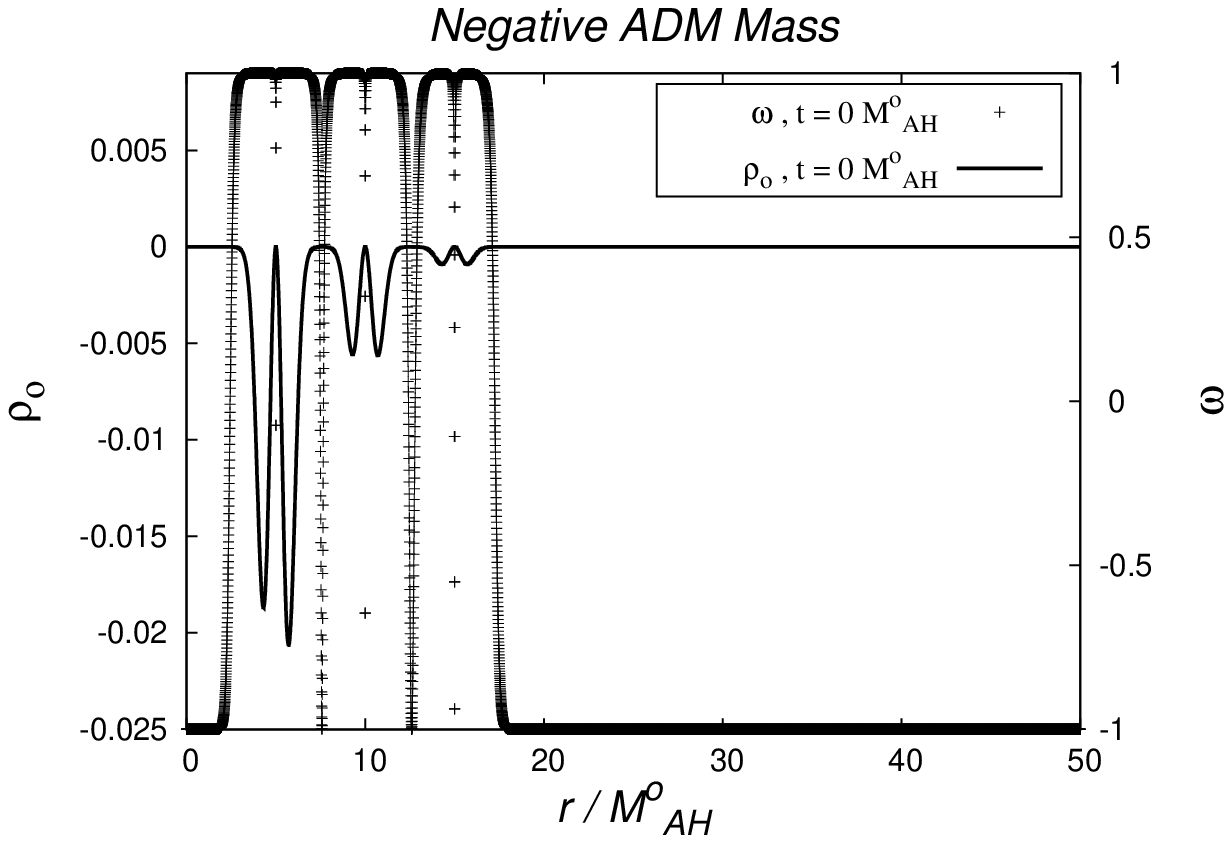}
\caption{In the top panels we show the initial profile of $\phi$ as function of the radial coordinate
(in units of the initial mass of the apparent horizon) for the exponential potential in the case of positive 
ADM mass (a single Gaussian) and negative ADM mass (train 
of three Gaussians); these profiles were used to solve the Hamiltonian constraint at initial time.
In the bottom panels we show the energy density of the scalar field $\rho=T_{\mu\nu}n^{\mu}n^{\nu}$ labeled with the left axes, and the equation of state $\omega=p/\rho$ 
labeled with the right axes as function of the radial coordinate (using the same units as before)
at initial time for two cases:  i) the initial scalar field profile corresponds to a Gaussian $\phi(r,t=0)=0.28 e^{-(r-5)^2}$ and exponential 
potential, and ii) the initial scalar field profile corresponds to the superposition of three gaussians $\phi(r,t=0)=0.25 e^{-(r-5)^2}+0.125 e^{-(r-10)^2}+0.05 
e^{-(r-15)^2}$, also for the exponential potential. At further values of time, the behavior of $\omega$ is the same, except that it appears shifted in the spatial 
direction. This behavior indicates that $\omega$ does not only depend highly on the space, but also on time.}
\label{fig:id}
\end{figure*}

In Fig. \ref{fig:id} we show a particular case of an initial scalar field profile. We can see that the solution of the Hamiltonian constraint implies already at initial time a negative energy density as expected from the violation of the null energy condition of this type of field. This also opens another possibility to be explored, that is, once we know that the scalar field contributes with a negative energy density it is also possible to construct initial configurations whose ADM mass is negative, that is, the contribution of the scalar field energy density to the space-time is bigger in absolute value than that due to the black hole's horizon.

On the other hand, we also show in Fig. \ref{fig:id} the relation $\omega = p/\rho$ where $p = T^{r}{}_{r}$, which clearly indicates that at local 
scales, the Lagrangian of the phantom scalar field does not behave at all as phantom matter since $p/\rho > -1$, and in fact runs from  cosmological constant $\omega=-1$ to stiff matter $\omega=1$.

In our study we set up two classes of initial data corresponding to positive ADM mass and negative ADM mass for all the potentials studied. In order to reduce the 
parameter space of our study, for each potential we fine-tune the initial scalar field profile so that the ADM mass  has the same value for all the potentials explored. 
The potential we used are: a) quadratic: $V(\phi) = V_0 \phi^2$, b) quartic: $V(\phi) = V_0 \phi^2 + V_1 \phi^4$ and c) exponential: $V(\phi) = V_0 e^{-\phi}$, where $V_0$ and $V_1$ are negative 
constant parameters. We also explored the potentials $V(\phi)=V_0(\cosh\phi-1)$ and $V_0=\sinh\phi$, however the behavior with these potentials is very similar to the 
obtained with the potentials mentioned before.

\begin{figure*}
\includegraphics[width=7cm]{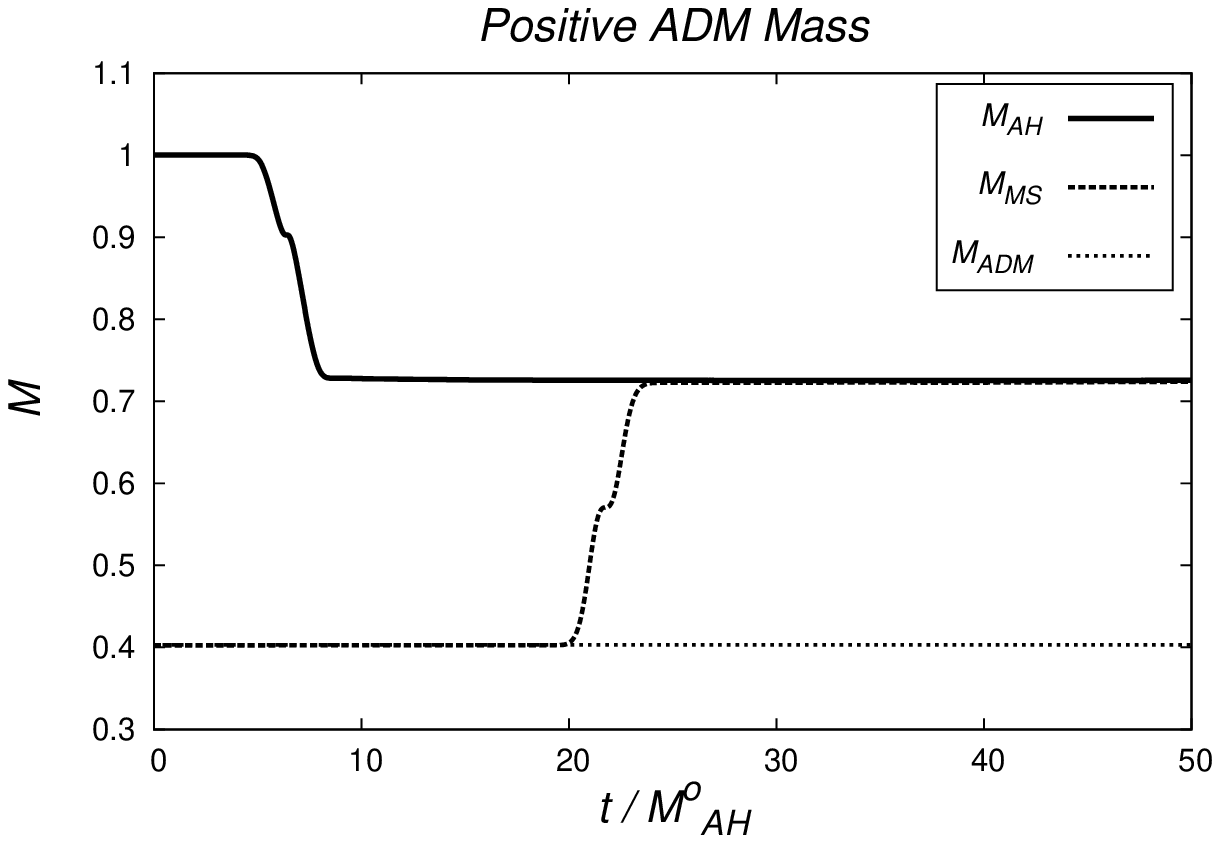}
\includegraphics[width=7cm]{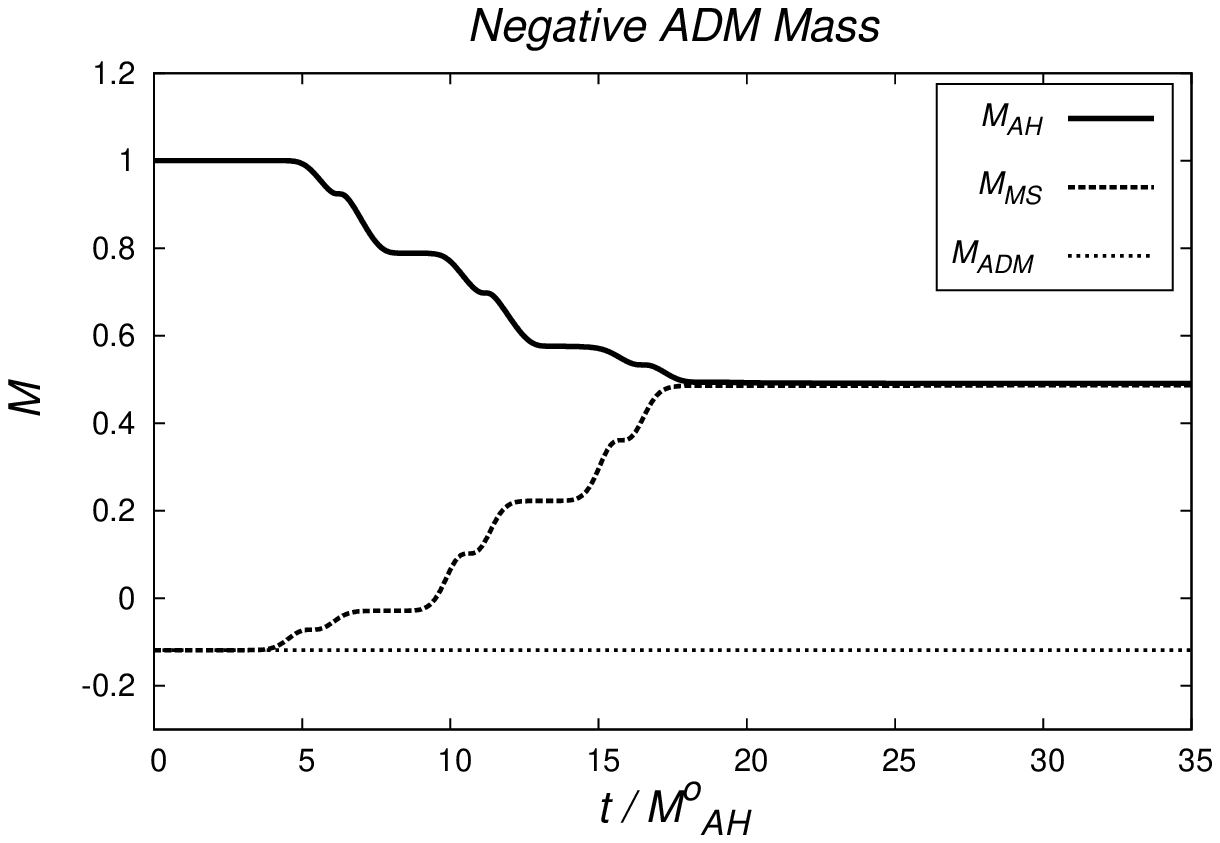}
\includegraphics[width=7cm]{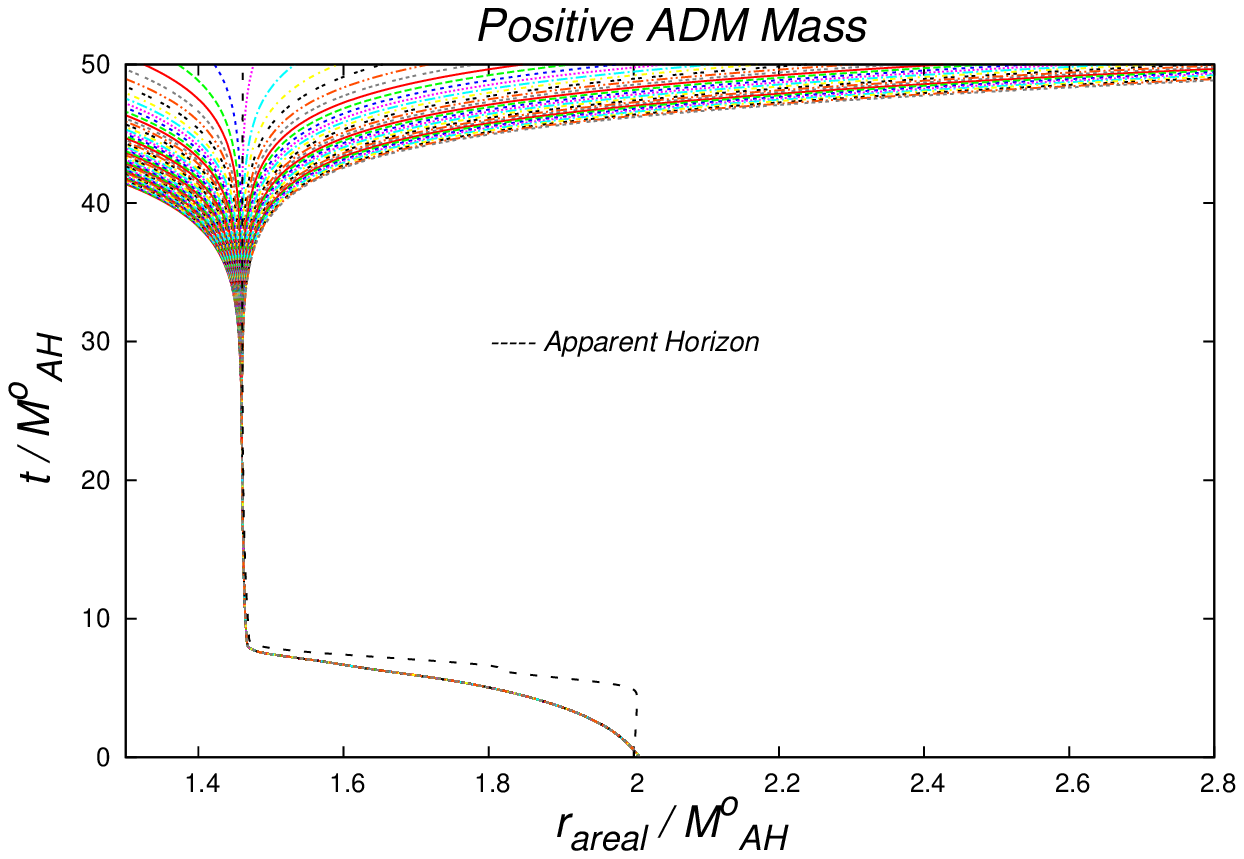}
\includegraphics[width=7cm]{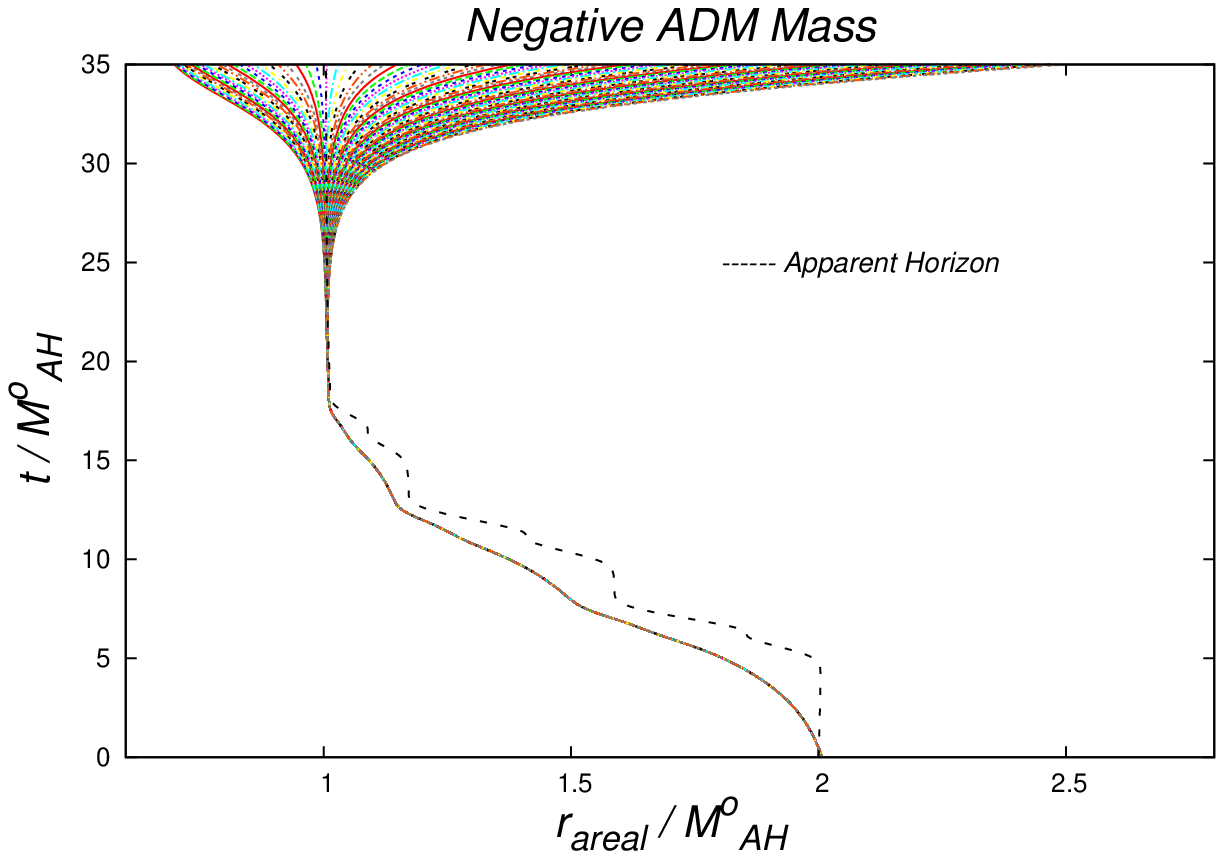}
\caption{In the top panels we present the behavior of the different masses versus the coordinate time for the exponential potential and both, positive and negative ADM 
mass cases. The solid line represent the mass of the apparent horizon, the bold dashed line is the Misner-Sharp mass and the thin dashed line is the ADM mass of the 
space-time. In the positive ADM mass the apparent horizon of the black hole reduces up to 30 percent of its initial mass whereas in the negative case it is reduced up 
to 50 percent. In the bottom panel we show a set of outgoing null rays that help estimating approximately the location of the event horizon. For comparison we also show 
the location of the apparent horizon: at early stages of the accretion the apparent horizon lies outside of the event horizon, which is not surprising given that the 
energy conditions are not fulfilled and we notice that after the system has stabilized the event horizon and the apparent horizon coincide.}
\label{fig:massADM}
\end{figure*}

In Fig. \ref{fig:massADM} we show the accretion of the scalar field for all the potentials, in particular, we show the mass of the apparent horizon and the Misner-Sharp 
masses. The finding is that despite of the potential used, the final mass of the apparent horizon is nearly the same in all cases, that is, the potential has no effect 
on the final state of the resulting black hole.

For both classes of configurations, positive and negative ADM mass, we found that the apparent horizon mass decreases and converges asymptotically to the Misner-Sharp 
mass. As a particular example, we show in Fig. \ref{fig:massADM} the behavior of the apparent horizon mass, the Misner-Sharp mass and the even horizon location for the 
exponential potential case.

In order to make sure that the decrease of the black hole is real, we fine-tuned a bundle of outgoing null rays for the positive and negative ADM mass cases, and 
estimated the location of the event horizon, since it is gauge invariant. We show the results in Fig. \ref{fig:massADM} where we present the location of the event 
horizon for the exponential potential case. The behavior shown in Fig. \ref{fig:massADM} is generic and applies to all the cases studied, that 
is, the event horizon -not only the apparent horizon- shrinks and is located approximately at the apparent horizon surfaces when the system has stabilized.

\begin{figure*}
\includegraphics[width=7cm]{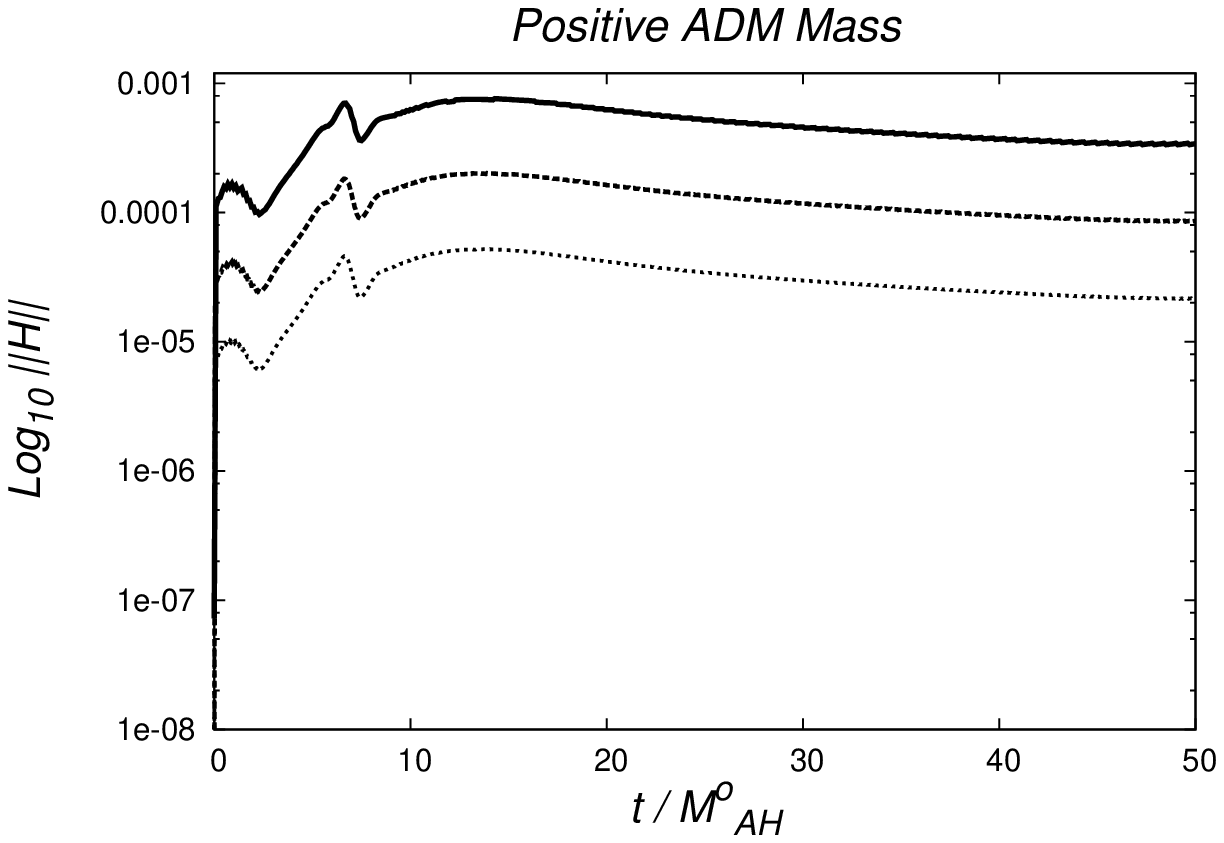}
\includegraphics[width=7cm]{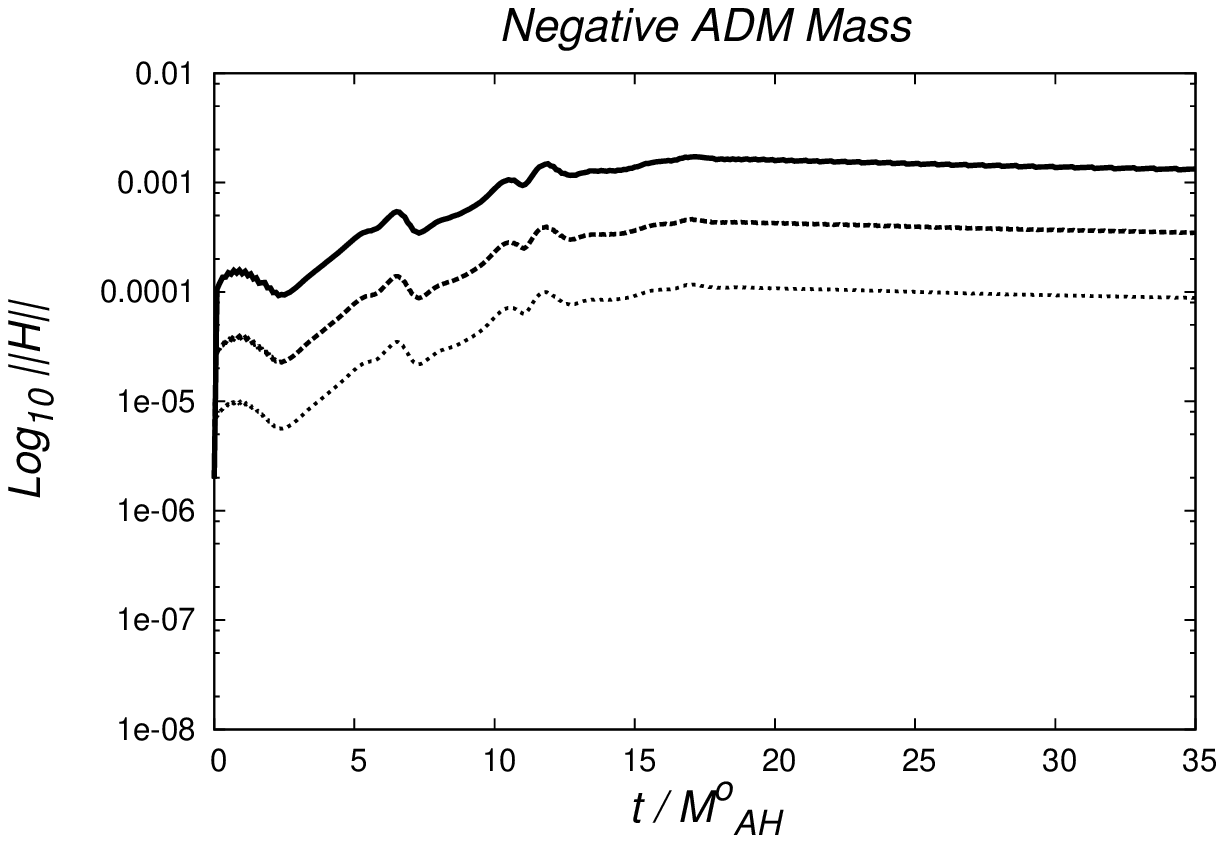}
\includegraphics[width=7cm]{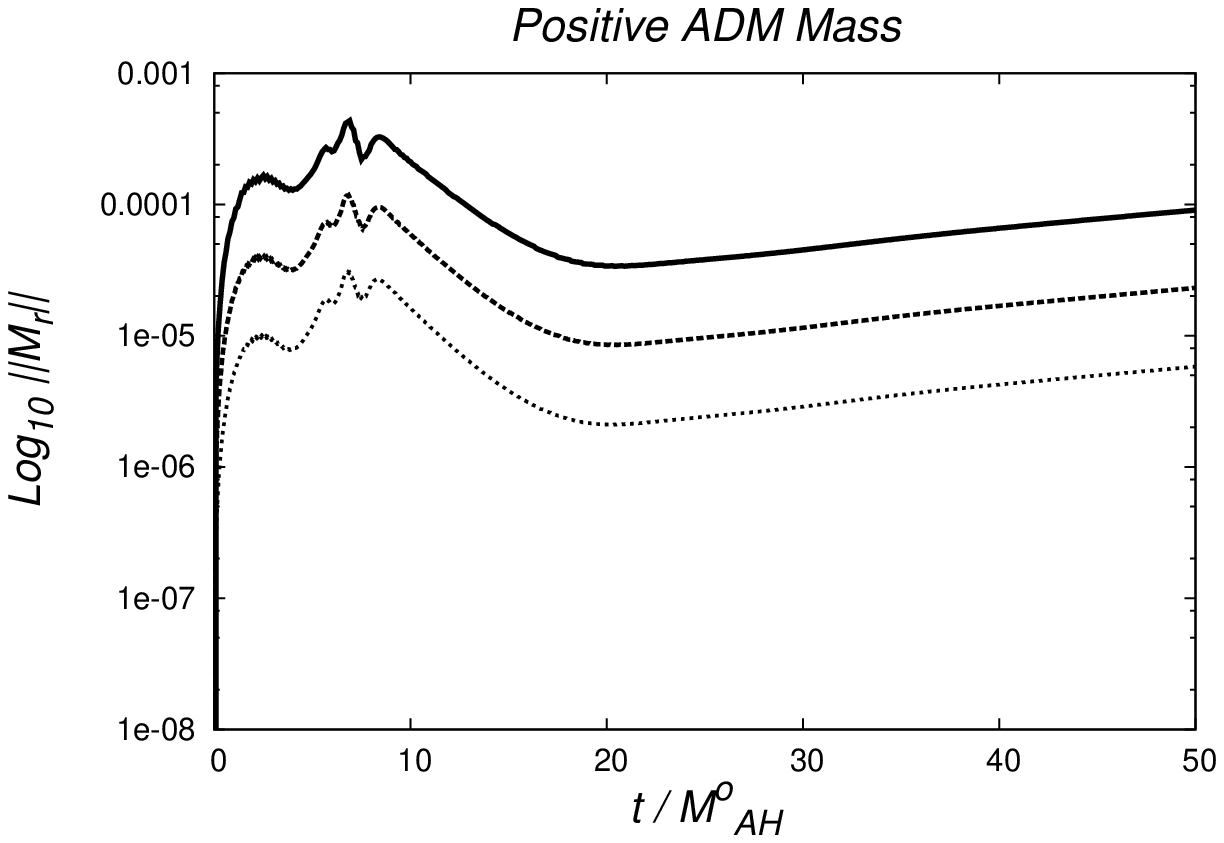}
\includegraphics[width=7cm]{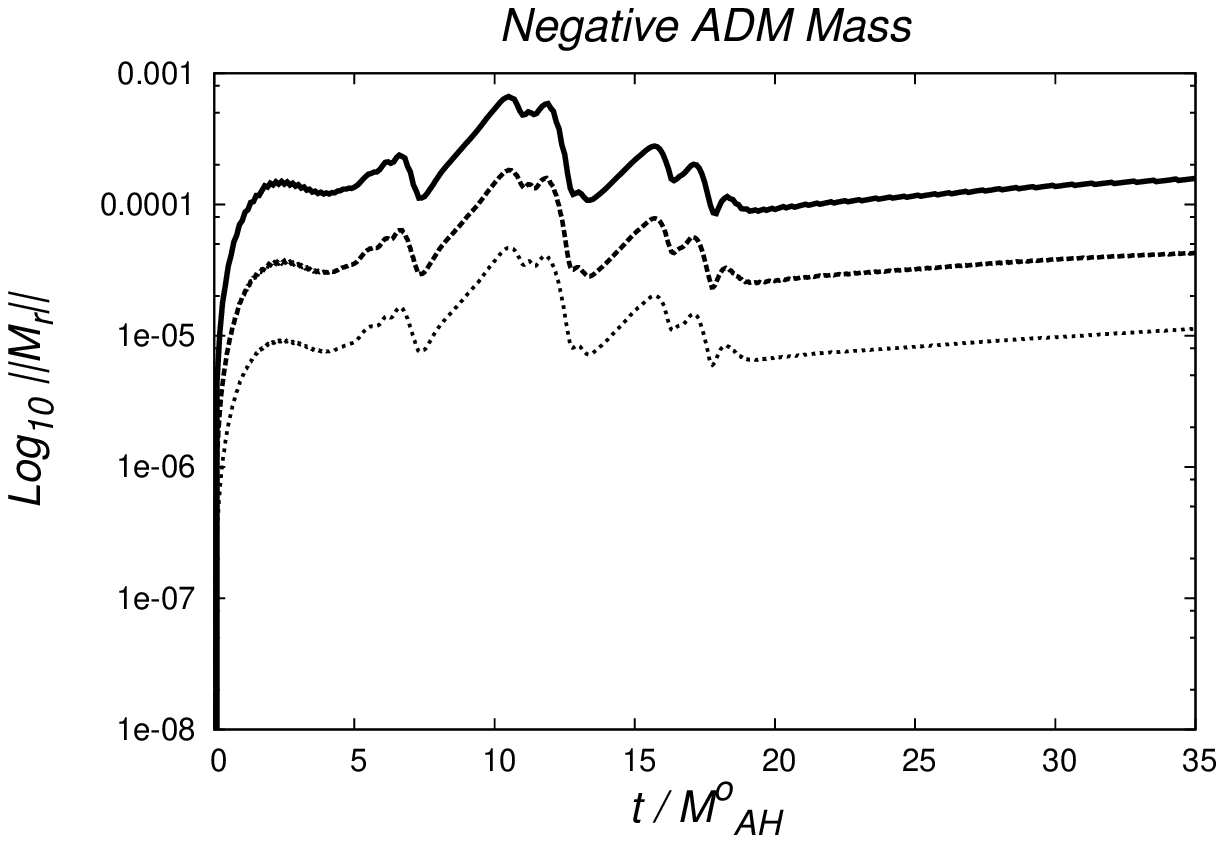}
\includegraphics[width=7cm]{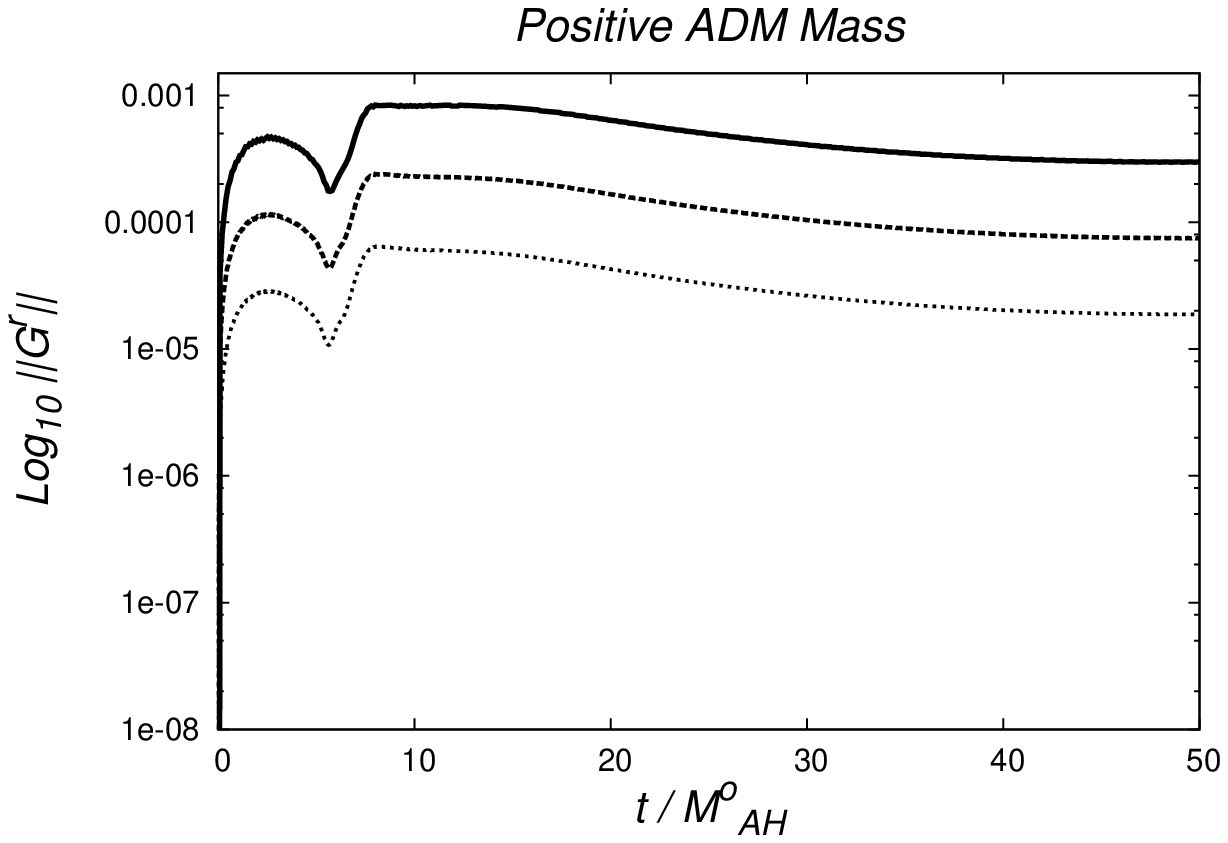}
\includegraphics[width=7cm]{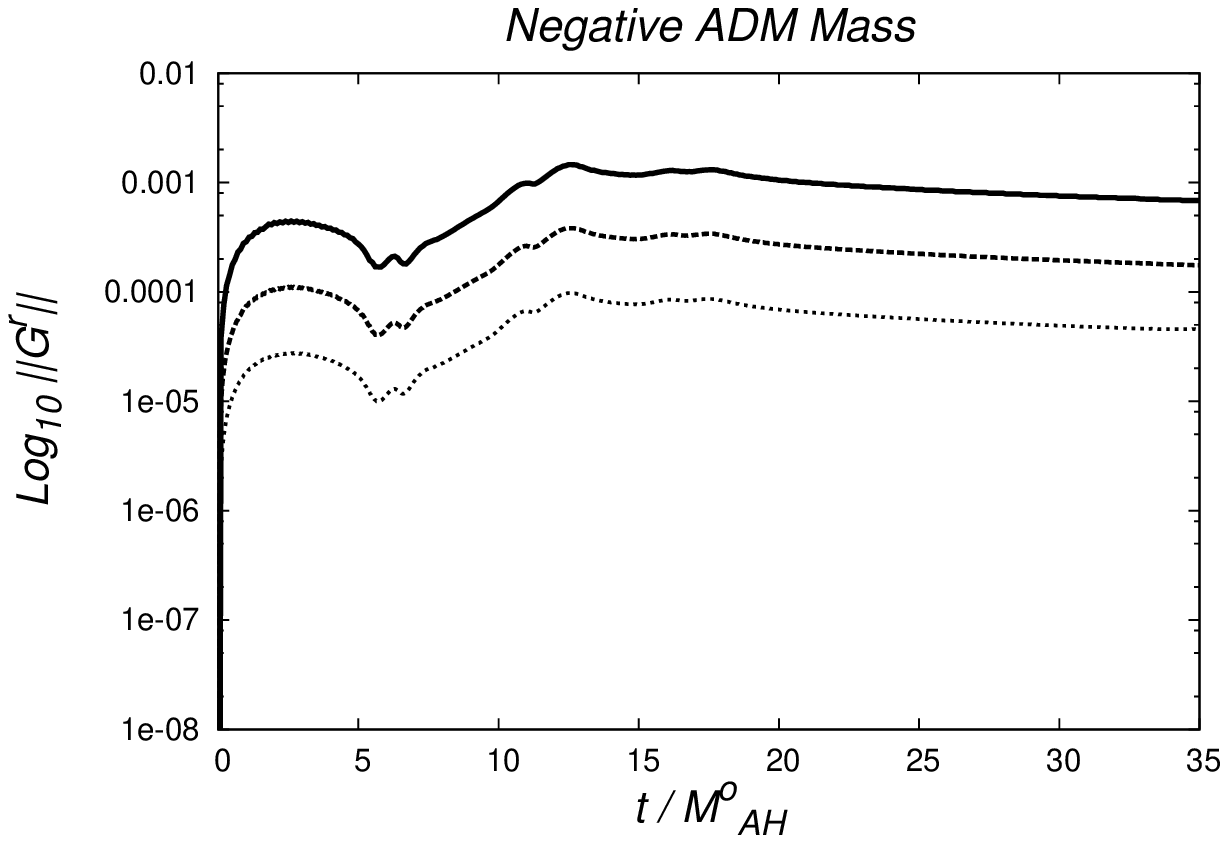}
\caption{We show the convergence of the $L_2$ norm of the violation of the Hamiltonian, momentum and $G^r$ constraints to zero for positive and negative ADM mass cases.}
\label{fig:pconver}
\end{figure*}

Finally, in order to validate our numerical results we show in Fig. 
\ref{fig:pconver} the convergence of the $L_2$ norm of the constraints to zero as the resolution is 
increased with fourth order for two representative cases with the 
exponential potential. 

\begin{figure*}
\includegraphics[width=7cm]{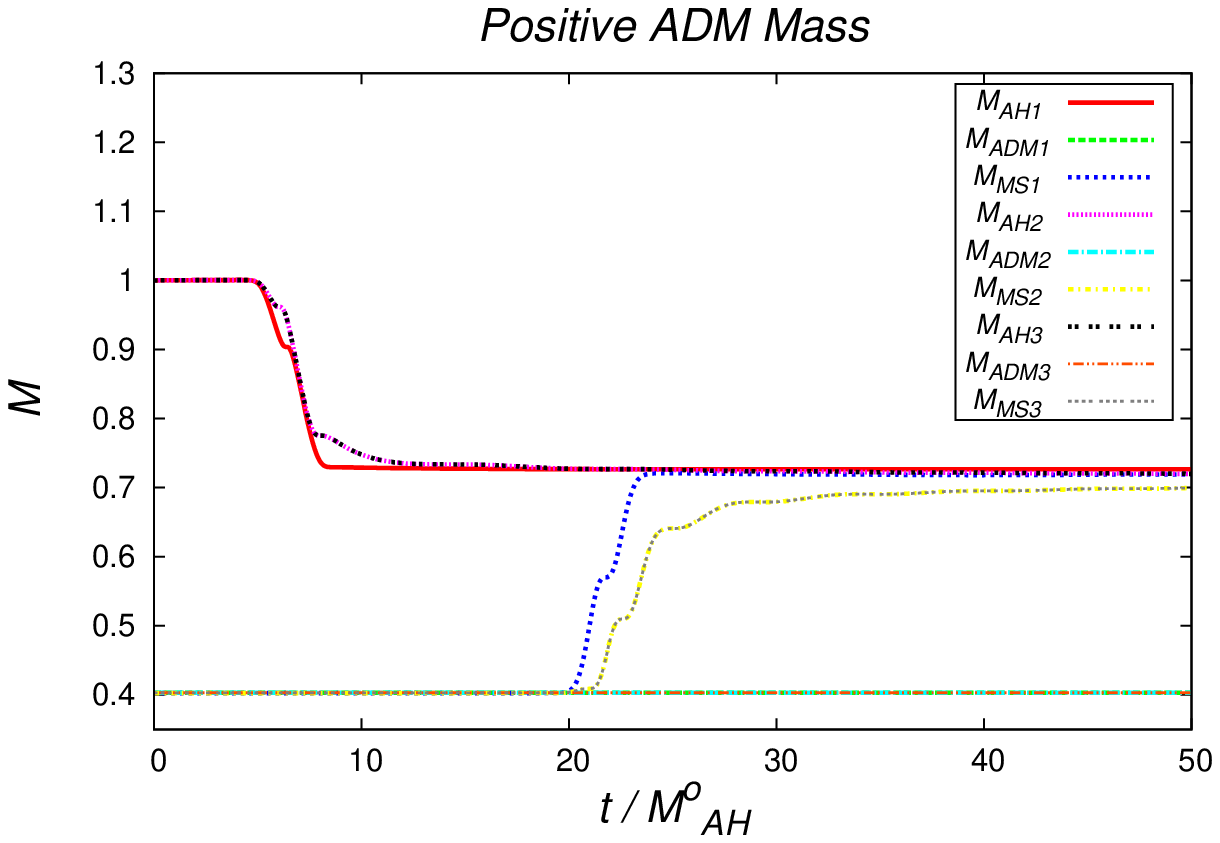}
\includegraphics[width=7cm]{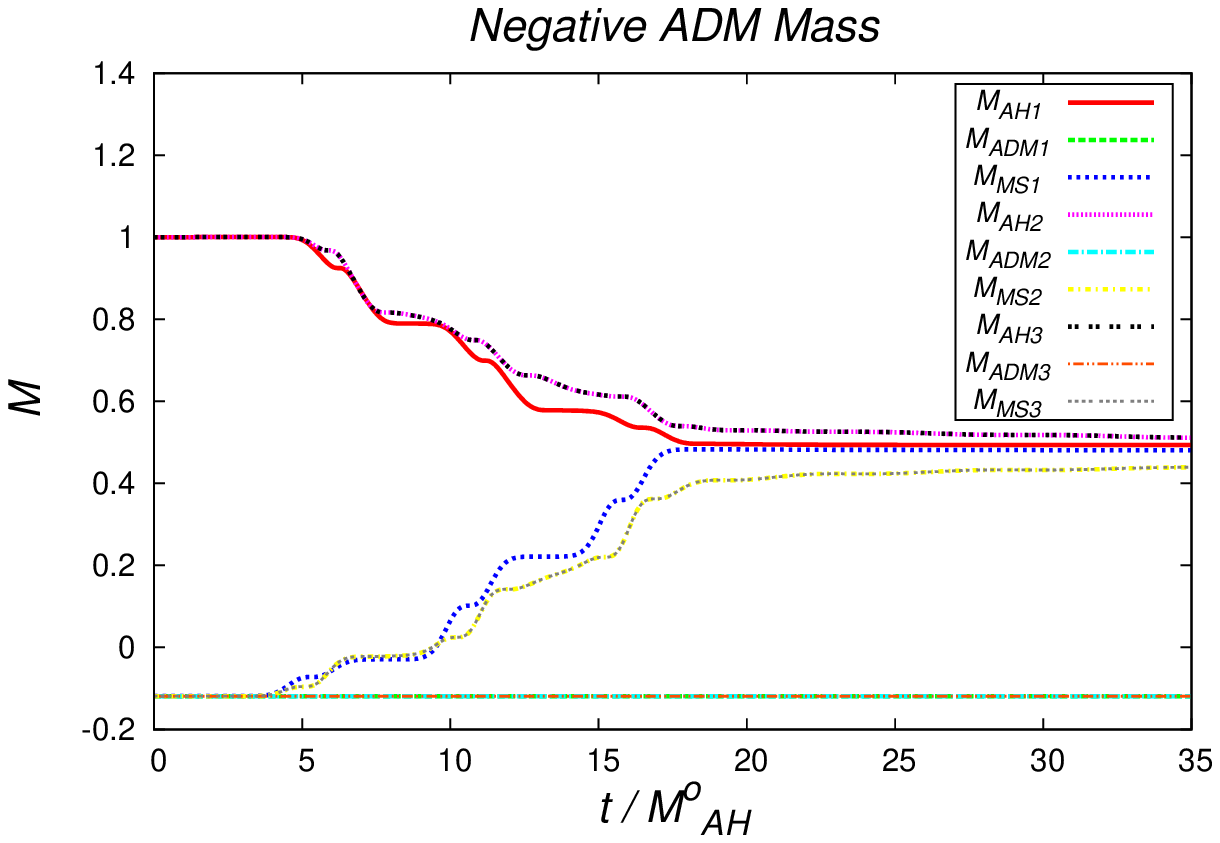}
\caption{In these plots, we show the behavior of the apparent horizon, Misner-Sharp and ADM masses with three self-interaction potentials:  
i) curves with subindex 1 correspond to the exponential potential. 
ii) curves with subindex 2 correspond to the quadratic potential. 
iii) Curves with subindex 3 correspond to the quartic potential. 
In the case of the exponential potential the decrease is faster than in the other cases, however the final mass of the horizon is not quite different considering that 
the scalar field profile is so free that the in-going profile depends on the scalar field potential.}
\label{fig:masscom}
\end{figure*}

In Fig. \ref{fig:masscom} we summarize the results related to the accretion mass rate and final mass of the apparent horizon for all various potential used. In each case 
we fine-tuned different scalar field profiles for the different 
potentials in order to use only one value of the ADM mass for the 
positive case and one value for the 
negative case.


\section{Conclusions}
\label{sec:conclusions}

We present the full non-linear spherical accretion of a phantom scalar field with various potentials into a black hole.

This accretion process is very efficient at reducing the black hole horizon area, which is a potentially important process that may impose restrictions on the existing black hole masses, or for instance primordial black holes.
In our simulations we were able to reduce the mass of the initial black hole up to 50$\%$. Attempts to reduce the area even further presents difficulties with our approach due to the steep gradients in the equations near the origin, and other techniques might be implemented in order to sort this problem out.

We show that the final mass of the black hole's horizon is nearly independent of the potential used. This results is physically interesting because the process of area reduction will apply independently of the cosmologically motivated potential used.

We show that even though the Lagrangian (\ref{eq:lagrangian}) of the scalar field corresponds to a phantom field, 
at local scales the radial equation of state is always $p/\rho \ge -1$, and it depends on space and time, with 
values ranging from a cosmological constant equation of state $\omega=-1$ up to a stiff mater $\omega =1$.

We want to stress the importance of the equation of state of the scalar field given by the Lagrangian (\ref{eq:lagrangian}), 
and point out that perhaps the big rip scenario would strongly depend on the homogeneity and isotropy of the scalar field. 
We have shown how much the equation of state can depend on space and time for the model of a scalar 
field, at least in the strong gravitational field regime.

The astrophysical possibility of this accretion process strongly depends on the chance of having a phantom scalar 
field located near a black hole with a rather sharp profile.


\section*{Acknowledgments}

This work was supported in part by grants 
CIC-UMSNH 4.9 and 4.23,
PROMEP UMICH-PTC-210, UMICH-CA-22 and UMICH-CA-22 network from SEP Mexico,
and 
CONACyT grant numbers 79601 and 106466.
The runs were carried our in the IFM cluster.



\begin{thebibliography}{99}
\bibitem{Corasaniti}
	P. S. Corasaniti et al., 
	Foundations of observing dark energy dynamics with the Wilkinson Microwave Anisotropy Probe.
	Phys. Rev. D {\bf 70}, 083006, 2004.

\bibitem{Babichev}
	Babichev E., Dokuchaev V., and Eroshenko Yu..
	Black hole mass decreasing due to phantom energy accretion. 
	Phys. Rev. Lett., 93, 021102 (2004).

\bibitem{Faraoni}
	Gao Ch., Chen X., Faraoni V., and Shen Y-G, 
	Does the mass of a black hole decrease due to the accretion of phantom energy?
	Phys. Rev. D 78, 024008 (2008).

\bibitem{Guzman}
        Gonz\'alez J.A., and Guzm\'an F.S.,
        Accretion of phantom scalar field into a black hole. 
        Phys. Rev. D {\bf 79}, 121501 (2009).

\bibitem{Brown}
	Brown J.D., 
	BSSN in spherical symmetry.
	Class. Quantum Grav., {\bf 25}, 205004 (2008).

\bibitem{Thornburg}
	Thornburg J.,
	A 3+1 computational scheme for dynamic spherically symmetric black hole space-times II: time 		evolution.
	Phys. Rev. D 59, 104007, (1999).

\bibitem{BSSN}
	M. Shibata, T. Nakamura,
	Evolution of three dimensional gravitational waves: harmonic slicing case. 
	Phys. Rev. D {\bf 52} 5428, 1995.
	T. W. Baumgarte, S. L. Shapiro,
	On the numerical integration of Einstein's field equations. 
	Phys. Rev. D {\bf 59} 024007,1999.

\bibitem{NASA} 
	John G. Baker, et al..
	Gravitational wave extraction from inspiraling configuration of merger black holes.
	Phys. Rev. Lett. 96, 111102 (2006).

\bibitem{Brownsville} 
	M. Campanelli, et al., 
	Accurate evolution of orbiting black hole binaries without excision. 
	Phys. Rev. Lett. 96, 111101 (2006).

\bibitem{Bruegmann} 
	S. Brandt, B. Bruegmann,
	A simple construction of initial data for multiple black holes. 
	Phys. Rev. Lett {\bf 78} 3606, 1997.

\bibitem{Brown2}
	Brown J.D., 2005,
	Conformal invariance and the conformal--traceless decomposition of the gravitational field. 
	Phys. Rev. D {\bf 71}, 104011.


\bibitem{excision}
	Brown J. D. et al., 
	Excision without excision: the relativistic turducken. 
	Phys. Rev. D {\bf 76}, 051503 (2007).




	
\end{thebibliography}
\end{document}